\documentclass[aps, prd, twocolumn,a4paper]{revtex4}
\usepackage{ulem}
\usepackage{color}
\usepackage{float}
\usepackage{graphicx}
\usepackage{epsfig}
\usepackage{epstopdf}
\usepackage{amsmath}
\usepackage{subfig}
\usepackage[dvipsnames]{xcolor}
\newcommand{\be}{\begin{equation}}
\newcommand{\ee}{\end{equation}}
\newcommand{\ba}{\begin{eqnarray}}
\newcommand{\ea}{\end{eqnarray}}

\usepackage{graphicx}
\usepackage{color}
\begin{document}
\title{Thermodynamics and relativistic kinetic theory for q-generalized Bose-Einstein and Fermi-Dirac systems}
\author{Sukanya Mitra}
\email{sukanyam@iitgn.ac.in}
\affiliation{Indian Institute of Technology Gandhinagar, Palaj, Gandhinagar-382355, Gujarat, India}

\begin{abstract}

The thermodynamics and covariant kinetic theory have been elaborately investigated in a non-extensive environment considering
the non-extensive generalization
of Bose-Einstein (BE) and Fermi-Dirac (FD) statistics. Starting with Tsallis' entropy formula, the fundamental principles of 
thermostatistics have been established for a grand canonical system having q-generalized BE/FD degrees of freedom. The many
particle kinetic theory has been set up in terms of the relativistic transport equation with q-generalized Uehling-Uhlenbeck collision
term. The conservation laws have been realized in terms of appropriate moments of the transport equation. The thermodynamic quantities 
have been obtained in weak non-extensive environment for a massive pion-nucleon and a massless quark-gluon
system with non-zero baryon chemical potential. In order to get an estimate of the impact of non-extensivity on the system dynamics, 
the q-modified Debye mass and hence the q-modified effective coupling have been estimated for a quark-gluon system.
\\
{\bf PACS}: 

\end{abstract}
\maketitle

\section{Introduction}

Boltzmann-Gibbs (B-G) statistics, has long been served as the founding structure of a wide range of physical systems, 
especially the ones containing a large number of particles, in the presence of short range correlations  
(exponentially decaying). Any understanding beyond such correlations where the memory effects are significant and 
non-Markovian processes are likely to occur, requires theoretical modelling of the system under consideration in 
terms of a generalized statistical approach, such that in an appropriate limiting case it yields the usual B-G statistics. 
In this context, Tsallis approach which is the non-extensive generalization of B-G statistics 
\cite{Tsallis1,Tsallis2,Tsallis3,Tsallis4,Tsallis5,Tsallis6,Tsallis7,Tsallis8,Abe1,Abe2,Martinez1,Martinez2},
could serve the purpose in developing the understanding of such systems with long range correlations.

In context of high energy collisions, the observables (such as particle spectra and transverse momentum fluctuations) in certain 
situations could be quantitatively explained better in terms of non-extensive statistics producing power law distributions. In
high energy experiments, the 
momentum distribution provided by B-G statistics being exponential in nature gives sensible description of particle production data 
only at low transverse momentum (below $p_{T}\sim 1 \textrm{GeV}$). However, for higher $p_{T}$ ranges, the particle spectra is observed to follow a 
power-law type tail. In a number of current literature, the requirement of a generalized statistics leading to a power law kind of 
distribution of emitted particles in high energy collisions, have been extensively studied \cite{Alberico1,Alberico2,Alberico3,Wilk1,Wilk2,Beck,Biro1}.
This could be a hint towards the presence of long range correlations among the particles within the system. In particular, for hot 
QCD systems produced at relativistic heavy ion collision experiments, this condition looks more feasible. Although in experimental
facilities like RHIC (Relativistic Heavy Ion Collider at BNL, USA) and LHC (Large hadron Collider at CERN, Geneva), due to large 
momentum transfer the interaction strength among the quarks and gluons becomes weaker by the virtue of asymptotic freedom, that 
can only be realized at temperatures much beyond the transition temperature $(T_c)$ from hadron to quark-gluon plasma (QGP) 
phase. Around $T_c$ the interaction is considerably large, giving rise to a strongly coupled QGP where the partonic degrees of freedom
could be deconfined just beyond their nucleonic volume. This phenomenon leads to long range entanglements and consequently the memory 
effects of microscopic interactions are significant. 

The two-particle, long-range correlations in the systems created in collider experiments already have been identified in a good deal 
of recent works where the correlation is quantified by the ``ridge structure'', observed in particle multiplicity distributions as a 
function of relative pseudo-rapidity \cite{CMS1,CMS2,CMS3,PHENIX,Li}. Clearly the B-G statistics which works wonderfully well for systems 
having close spatial connections and short ranged time connections, fail to describe such exotic systems at a quantitative level.
Further the ``Hypothesis of molecular chaos'' from Boltzmann kinetic theory assuming the interacting particles to be uncorrelated,
does not hold in such situation and non-Markovian processes become relevant where the consequences of particle
collisions are no more independent of its past interaction memory. The non-Markovian processes with long-range correlations
have been precisely studied within the frame work of the Tsallis statistics in \cite{Caceres}. The complexity of the system where the 
space-time and consequently the phase space are non-fractal, proves to be beyond the scope of standard B-G statistics and hence
for a non-trivial system like QGP the non-extensive generalization becomes quite relevant.

In the context of describing the strongly interacting system created in heavy ion collisions, already Tsallis statistics have been applied
to observe different concerned phenomena. Apart of describing the transverse momentum spectra as already mentioned, a large number 
of observables and system properties have been analyzed in the light of non-extensive dynamics. The relativistic equation of state 
of hadronic matter and quark-gluon plasma at finite temperature and baryon density have been investigated in the framework of the 
non-extensive statistical mechanics in \cite{Gervino,Pereira,Lavagno0}. The non-extensive statistical effects in the hadron to quark-gluon phase transition
have been studied in the works \cite{Lavagno1,Lavagno2}. The strangeness production has been mentioned under the same statistical
set up in \cite{Lavagno3}. Local deconfinement in relativistic systems including strong coupling diffusion and memory effects
have been discussed in \cite{Wolschin1,Wolschin2}. The kinetic freeze-out temperature and radial flow velocity have been
extracted from an improved Tsallis distribution in \cite{Lacey}. A modified Hagedorn formula and also limiting temperature concept
of a parton gas with power-law tailed distribution presented respectively in \cite{Wilk3} and \cite{Biro2}. In \cite{Wilk4,Wilk5,Tawfik}, 
effects of non-extensive statistics have been manifested in terms of the fluctuation of system variables such as 
temperature and chemical fluctuations in high-energy nuclear collisions. Nonextensive Boltzmann Equation and associated 
hadronization of quark matter have been discussed in \cite{Biro3}. Recently a considerable amount of work has been carried out regarding the 
bulk observables of ultra-relativistic heavy ion collisions, such as the transverse momentum distribution, radial flow, 
evolution of fluctuation, nuclear modification factor, speed of sound etc. under the Tsallis' generalization of B-G statistics
\cite{Cleymans3,Cleymans4,Cleymans5,Trambak1,Trambak2,Trambak3,Trambak4,Parvan,Khuntia1,Khuntia2,Cleymans1,Cleymans2}.

In view of the facts mentioned above, the Tsallis statistics has a vast applicability in strongly interacting systems
related to heavy ion physics. Thus, the need for setting up a complete model involving relativistic kinetic theory and fluid
dynamics under this generalized statistical scheme is desirable. In this context, there are a few attempts which provides 
macroscopic thermodynamic quantities such as particle number density, energy density,
pressure, equation of state and transport parameters, starting from a relativistic microscopic theory \cite{Lavagno4,Biro-Molner1,Biro-Molner2,Wilk6}.
In most of these studies the nonextensive generalization has been made over the Boltzmann distribution function while describing
the single particle momentum distribution, but the quantum statistical effects of Bose-Einstein (BE) and Fermi-Dirac (FD) distributions are missing.
The quantum statistical factors (Bose enhancement for BE and Pauli blocking for FD systems respectively) essential to describe a QCD 
(quantum chromodynamics) system, are not being included in the phase space integral of the collision term while setting up the relativistic
transport equation. This fact sets the motivation for  the present investigations. 

In this work a complete thermostatistical model for a grand canonical system under nonextensive environment and  
hence a relativistic kinetic theory for a many particle system including B-E and F-D distributions for individual
species of particles have been formulated in detail. The quantum statistical factors in the momentum distribution of final state particles 
introduced by Uehling and Uhlenbeck in semiclassical transport theory, have been carefully included while developing the q-generalized
theory. Under the constructed formalism the thermodynamic macroscopic state variables and effective coupling for hot QCD system in non-extensive
environment have been estimated.

The manuscript is organized as follows. Section-II deals with the formalism part where firstly the non-extensive thermostatistics for a grand canonical 
ensemble and then the non-extensive relativistic kinetic theory with quantum statistical effects have been discussed in subsections A and B respectively.
In subsection C the analytical expressions of thermodynamic quantities have been obtained and finally subsection D is contributed to
estimate the effective coupling for a q-generalized QCD system. 
Section-III contains the results displaying the temperature dependence of the evaluated quantities with finite baryon chemical potential,
and the relevant discussions as well.
The manuscript has been completed incorporating the concluding remarks and possible outlooks of the present work in section-IV.

\section{Formalisms} 

Here, the salient features of the non-extensive thermostatistics for a grand canonical ensemble and the relativistic kinetic theory
for a multi-component system with constituents obeying BE/FD distributions have been discussed. The 
entropy maximization technique in the first case using the method of Lagrange's undetermined multipliers and in the second 
case by applying the laws of summation invariants have been shown to obtain the identical expressions of single particle BE/FD distribution 
function in a non-extensive environment. This consistency provides the ground for the microscopic definitions of the generalized entropy 
and collision integral including the quantum statistical effects of Bose enhancement and Pauli blocking. Afterwards, the distribution functions 
are employed to obtain a number of thermodynamic quantities essential to specify the macroscopic properties of a system. In the present work these 
quantities have been estimated for a massive pion-nucleon and a massless quark-gluon system with finite baryon chemical potentials. Finally the 
modification of QCD coupling describing the interaction dynamics of the system under the effects of non-extensivity has also been estimated in 
order to include the long range interaction measures.

\subsection{Non-extensive thermostatistics for a grand canonical ensemble}

To start with, non-extensive generalization of entropy proposed by Tsallis \cite{Tsallis1,Tsallis2,Tsallis3,Tsallis4,Tsallis5,Tsallis6,Tsallis7,Tsallis8}
is given below,
\begin{equation}
 S_{q}=k\frac{1-\sum_{i=1}^{W}p_{i}^q}{q-1}, ~~~~~[q\in \textrm{R}].
 \label{entropy_T1}
\end{equation}
Here $k$ is a positive constant. $p_{i}$ is the probability associated with the $i^{th}$ state and $W\in{\textrm{N}}$ 
is the total number of possible microscopic configurations of the system following the norm condition,
\begin{equation}
\sum_{i=1}^{W}p_{i}=1~.
\label{norm}
\end{equation}
It is quite straight
forward to realize for the limiting condition of the entropic index $q\rightarrow 1$, Eq.(\ref{entropy_T1}) reduces to the 
well known form of Boltzmann-Gibbs entropy namely $ \lim_{q\to 1} S_{q}=-k\sum_{i=1}^{W}p_{i}\textrm{ln} {p_{i}} .$
Eq.(\ref{entropy_T1}) can alternately also be defined with the help of generalized differential operator as,

\begin{equation}
S_q=-k\bigg\{D_{q}\sum_{i=1}^{W}p_{i}^{\alpha}\bigg\}_{\alpha=1},
\end{equation}
following the operator definition $D_q f(x)\equiv \frac{f(qx)-f(x)}{qx-x}$ which for $q\rightarrow 1$ reduces to $\frac{df(x)}{dx}$.
One important signature of the entropy definition given in (\ref{entropy_T1}) is their pseudo additive rule for a combined system
consisting of two individual systems $A$ and $B$,

\begin{equation}
 \frac{S_{q}(A+B)}{k}=\frac{S_{q}(A)}{k}+\frac{S_{q}(B)}{k}+(1-q)\frac{S_{q}(A)}{k} \frac{S_{q}(B)}{k}~.
\end{equation}

Now in order to obtain an equilibrium probability distribution in a grand canonical system, we need to extremize $S_{q}$ in presence 
of a set of appropriate constraints regarding the choice of internal energy and particle number \cite{Biiyiikkd}. So along with
the entropy definition from (\ref{entropy_T1}) and norm constraint from (\ref{norm}), the choice of internal energy and particle number 
is addressed from the Tsallis original third choice of energy constraint for a canonical system \cite{Tsallis2} , extending currently for 
a grand canonical one by the two following equations,

\begin{eqnarray}
 \frac{\sum_{i=1}^{W}p_{i}^{q}\epsilon_{i}}{\sum_{i=1}^{W}p_{i}^{q}}=\bar{E}~,
 \label{encon}\\
 \frac{\sum_{i=1}^{W}p_{i}^{q}n_{i}}{\sum_{i=1}^{W}p_{i}^{q}}=\bar{N}~.
 \label{pncon}
\end{eqnarray}
Here $i$ labels the possible quantum states of the whole system where $\bar{E}$ and $\bar{N}$ are the average energy and average number 
of particles of the same. The best known way to get a solution of this variational problem is using the method of Lagrange's undetermined
multipliers, for which the following expression is needed to be extremized,

\begin{equation}
 Q=\frac{S_q}{k}+\alpha\sum_{i=1}^{W}p_{i}-\beta \frac{\sum_{i=1}^{W}p_{i}^{q}\epsilon_{i}}{\sum_{i=1}^{W}p_{i}^{q}}
                                          -\gamma \frac{\sum_{i=1}^{W}p_{i}^{q}n_{i}}{\sum_{i=1}^{W}p_{i}^{q}},
\end{equation}
where $\alpha$, $\beta$ and $\gamma$ are the Lagrange's undetermined multipliers. Executing extremization and following
the standard thermodynamic definition such as $\beta=\frac{1}{kT}$ and $\gamma=-\beta\mu$ with $T$ as the system
temperature and $\mu$ as the chemical potential for each particle, we finally achieve the probability distribution
of a grand canonical system in terms of fundamental thermodynamic quantities,

\begin{equation}
  p_{i}=\frac{1}{Z_q}\bigg[ 1- (1-q)\frac{\{\frac{1}{T}(\epsilon_i-\bar{E})-\frac{\mu}{T}(n_i-\bar{N})\}}{\{\sum_{i=}^{W}p_i^q\}}\bigg]^\frac{1}{1-q},
  \label{prob-gn}  
\end{equation}
with
\begin{equation}
 Z_q=\sum_{i=1}^{W}\bigg[ 1- (1-q)\frac{\{\frac{1}{T}(\epsilon_i-\bar{E})-\frac{\mu}{T}(n_i-\bar{N})\}}{\{\sum_{i=}^{W}p_i^q\}}\bigg]^\frac{1}{1-q},
 \label{Z-gn}
\end{equation}
as the q-generalized grand canonical partition function.
From Eq.(\ref{prob-gn}) and (\ref{Z-gn}) a very useful identity can be achieved which is given below,
\begin{equation}
 Z_{q}^{1-q}=\sum_{i=1}^{W}p_{i}^{q}~.
\end{equation}
To get the expressions compactified it is now the time to define the q-generalized exponential and logarithmic
functions,
\begin{eqnarray}
 exp_{q}(x)=&&\{1+(1-q)x\}^\frac{1}{1-q}~,
 \label{exp-q}\\
 ln_{q}x=&&\frac{x^{1-q}-1}{1-q}~,
 \label{ln-q}
\end{eqnarray}
which reduces to the conventional exponentials and logarithms as $q\rightarrow1$.
With help of the q-exponential function the probability distribution as well as the partition function from Eq.(\ref{prob-gn})
and (\ref{Z-gn}) can be redefined as,
\begin{eqnarray}
 p_i=\frac{1}{Z_q} exp_{q}\bigg[ \frac{-\big\{\frac{1}{T}(\epsilon_i-\bar{E})-\tilde{\mu}(n_i-\bar{N})\big\}}{\big\{\sum_{i=1}^{W}p_{i}^{q}\big\}} \bigg]~,
 \label{prob-gn1}\\
 Z_{q}=\sum_{i=1}^{W} exp_{q}\bigg[ \frac{-\big\{\frac{1}{T}(\epsilon_i-\bar{E})-\tilde{\mu}(n_i-\bar{N})\big\}}{\big\{\sum_{i=1}^{W}p_{i}^{q}\big\}} \bigg]~,
 \label{Z-gn1}
\end{eqnarray}
with $\tilde{\mu}=\mu/T$.
The definition of temperature and chemical potential can be derived in terms of the macroscopic quantities as the following,
\begin{eqnarray}
 \frac{\partial S_q}{\partial \bar{E}}=\frac{1}{T}~,\\
 \frac{\partial S_q}{\partial \bar{N}}=-\frac{\mu}{T}~
\end{eqnarray}
which is consistent with situation for $q=1$.

Now the choice of energy and particle number constraints from Eq.(\ref{encon}) and (\ref{pncon}) respectively needs some discussions
offered below.
There are a number of reasons that this choice is an unique one which is free from the unfamiliar consequences with respect to the 
other choices proposed \cite{Tsallis2}. 
First, it's invariance under the uniform translation of energy and particle number spectrum ($\{\epsilon_i\}$ and $\{n_i\}$ respectively)
makes the thermostatistical quantities independent of the choice of origin of energy and particle number densities. Secondly, the
normalization condition is carefully preserved in this choice ($\ll 1 \gg_q=1$, where $\ll O_i \gg_q\equiv \frac{\sum_{i=1}^{W}p_{i}^{q}O_{i}}{\sum_{i=1}^{W}p_{i}^{q}}$).
Finally, the most important one is that, it preserves the additive property of generalized internal energy ($\bar{E}(A+B)=\bar{E}(A)+\bar{E}(B)$) in
the exact same form of standard thermodynamics ($q$=1). In other words, the microscopic energy conservation is retained macroscopically as well.
This is an extremely crucial property in order to describe the dynamics of the system.

Next, it can be trivially shown that Eq.(\ref{prob-gn1}) and (\ref{Z-gn1}) can be presented by a set of simpler expressions in terms of renormalized
temperature and chemical potential as follows,

\begin{eqnarray}
 p_i=\frac{1}{Z_q} exp_{q}\bigg[ -\big\{\frac{\epsilon_i}{T'}-\tilde{\mu}'n_i\big\}  \bigg]~,
 \label{prob-gn2}\\
 Z_{q}=\sum_{i=1}^{W} exp_{q}\bigg[ -\big\{\frac{\epsilon_i}{T'}-\tilde{\mu}'n_i\big\} \bigg]~,
 \label{Z-gn2}
\end{eqnarray}
 
with
\begin{eqnarray}
 T'=TZ_{q}^{1-q}+(1-q)\bar{E}-\mu(1-q)\bar{N}~,
 \label{T-renorm}\\
 \tilde{\mu}'=\tilde{\mu}\frac{1}{Z_q^{1-q}+\frac{1}{T}(1-q)\bar{E}-\tilde{\mu}(1-q)\bar{N}}~.
 \label{mu-renorm}
\end{eqnarray}
Systems using some arbitrary finite temperature rather than specific temperature dependence of
involved thermostatistical quantities (and so for chemical potential), can conveniently use the definitions provided by Eq.(\ref{prob-gn2}) and
(\ref{Z-gn2}). The redefined expressions of temperature ($T'$) and chemical potential ($\mu'$) will be
denoted by the arbitrary temperature ($T$) and chemical potential ($\mu$) hereafter.

We will now proceed to obtain the single particle distribution functions for the bosons and fermions.
The quantum mechanical state of an entire system is uniquely specified only when the occupation 
numbers of the single particle states are explicitly provided. Thus the total energy and particle number of the system when it is
in the state $i$, with $n_1$ number of particles in state $k=1$, $n_2$ number of particles in state $k=2$ and so on,
is given by,
\begin{eqnarray}
 \epsilon_i=\sum_{k}n_k e_k~,~~~~~
 n_i=\sum_{k}n_k~.
\end{eqnarray}
The sum extends over all the possible states of a particle and $e_k$ is the energy of a particle in a single
particle state $k$.
Hence in terms of the single particle occupation states the expression for partition function from (\ref{Z-gn2}) becomes \cite{Biiyiikkd},
\begin{equation}
 Z_q=\prod_{k=1}^{\infty}\sum_{n_k=0}^{\infty} \big[1-(1-q)\frac{1}{T}(e_k-\mu)n_k\big]^\frac{1}{1-q}
 \label{Z-gn3}
\end{equation}
From Eq.(\ref{Z-gn3}) it is trivial to obtain from the standard technology of statistical mechanics, the single
particle distribution function for Bose-Einstein and Fermi-Dirac systems respectively as the following,

\begin{equation}
 f_{q}=\frac{1}{\big[1-(1-q)\{\frac{e}{T}-\frac{\mu}{T}\}\big]^{\frac{1}{q-1}}\mp 1}~.
 \label{dist1}
\end{equation}
Here $e$ and $\mu$ denote the single particle energy and chemical potential along with $T$ as the bulk
temperature of the system under consideration. Eq.(\ref{dist1}) gives us the expression for q-generalized
Bose-Einstein (BE) and Fermi-Dirac(FD) single particle distribution function in a non-extensive
environment.

\subsection{Non-extensive relativistic kinetic theory with quantum statistical effects}

In this section, the basic macroscopic thermodynamic variables will be defined in the frame work of a relativistic
kinetic theory in a non-extensive environment. For this purpose we first need to provide the microscopic definition for the 
q-generalized entropy and then set up a kinetic or transport equation describing the space-time behavior of one
particle distribution function.

In terms of the single particle distribution function the entropy 4-current is defined in the following integral form 
for a multicomponent system with $N$ number of species \cite{Cleymans1,Cleymans2},
\begin{eqnarray}
 S_{q}^{\mu}(x,q)&&=-\sum_{k=1}^{N}\int\frac{d^3p_k}{(2\pi)^3 p_k^0}p_{k}^{\mu}\nonumber\\
 &&\times \bigg\{\big(f_{q}^{k}\big)^q ln_q f_{q}^{k} +\frac{1}{z}\big(1-zf_{q}^k\big)^{q}ln_q \big(1-zf_{q}^{k}\big)\bigg\}~,\nonumber\\
 \label{ent-1}
\end{eqnarray}
with $z=1$ for FD and $z=-1$ for BE case. For $z\rightarrow 0$, Eq.(\ref{ent-1}) reduces to q-generalized
Boltzmann entropy, $S_q^{\mu}(x,q)=-\sum_{k=1}^{N}\int\frac{d^3p_k}{(2\pi)^3 p^0_k}p_{k}^{\mu}\big(f^k_q\big)^q\big\{ln_qf^k_q-1\big\}$. 
$f_q^k(x,p_k,q)$ is the notation
for the single particle distribution function belonging to $k^{th}$ species in a non-extensive environment depending 
upon the particle 4-momentum $p^{\mu}_{k}$, space-time coordinate $x$ and the entropic index $q$. The $q^{th}$ power
over particle distribution function defining the thermodynamic quantities is justified from the discussions
of the last section.

Next the relativistic transport equation for a non-extensive system is presented where the distribution function
belonging to each component of the system satisfy the equation of motion in the following manner,

\begin{equation}
p_k^{\mu}\partial_{\mu}\big(f_q^k\big)^q=\sum_{l=1}^{N}C^{kl}_{q}~~~~~[k=1,\cdots\cdots,N]. 
\label{releq}
\end{equation}

The detailed q-generalized collision term $C_q^{kl}$ for a BE-FD system including the quantum statistical 
effects (Uehling-Uhlenbeck terms \cite{UU}), describing the binary collision $k+l\rightarrow i+j$, is presented below,

\begin{eqnarray}
C^{kl}_{q}=&& \frac{1}{2} \sum_{i,j=1}^{N}\int \frac{d^3 p_l}{(2\pi)^3 p_l^0} \int \frac{d^3 p_i}{(2\pi)^3 p_i^0} \int \frac{d^3 p_j}{(2\pi)^3 p_j^0}\nonumber\\
&&\times\bigg\{h_q\big[f_q^i,f_{q}^{j},(1\pm f_q^{k}),(1\pm f_{q}^{l})\big]W_{i,j|k,l}\nonumber\\
&&-h_q\big[f_q^k,f_{q}^l,(1\pm f_q^i),(1\pm f_{q}^j)\big]W_{k,l|i,j}\bigg\},
\label{coll}
\end{eqnarray}
where the $h_q$ factors are defined in terms of the particle distribution functions in the following manner,

\begin{eqnarray}
 h_q \big[f_q^k,f_{q}^l,(1\pm f_q^i),(1\pm f_{q}^j)\big]=exp_{q}\bigg[ln_qf_q^k+ln_qf_{q}^l\nonumber\\
 +\frac{(1\pm f_q^i)^{q-1}}{(f_q^i)^{q-1}}ln_q(1\pm f_q^i) +\frac{(1\pm f_{q}^j)^{q-1}}{(f_{q}^j)^{q-1}}ln_q(1\pm f_{q}^j)\bigg].
 \label{hq}
\end{eqnarray}
Here $W_{k,l|i,j}$ is the interaction rate for the binary collision process $k+l\rightarrow i+j$. The $1/2$ factor 
in the right hand side of Eq.(\ref{coll}) takes care of the indistinguishability
of the final state particles if their momenta are exchanged from $p^i,p^j$ to $p^j,p^i$.

Now an important remark has to be made
at this point. In the standard Boltzmann transport equation, the ``Hypothesis of molecular chaos'' or ``Stosszahlansatz'' postulates
that in binary collisions the colliding particles are uncorrelated. Which means any correlation present in an early time must be
ignored. Thus in Bolzmann transport equation the statistical assumption about the number of binary collisions occurring is
proportional to the simple product of the distribution functions of colliding particles along the quantum corrections of the 
final state distribution functions multiplied with the interaction rate \cite{Degroot}. However with systems where long range correlations are 
present, the memory effects are significant which lead to non-Markovian processes. Thus the q-generalized complex looking behavior
of the transport equation expressed by Eq.(\ref{releq}), (\ref{coll}) and (\ref{hq}) is explained in the present case.

From the expression for entropy 4-current (\ref{ent-1}), we can obtain the expression for local entropy production as the following,

\begin{equation}
 \sigma(x,q)=
 \partial_{\mu}S_{q}^{\mu}(x,q)~,
\end{equation}

which for the present case turns out to be,

\begin{equation}
 \sigma=-\sum_{k=1}^{N}\int\frac{d^3p_k}{(2\pi)^3p_k^0}p_k^{\mu}\partial_{\mu}\big(f_q^k\big)^q\big\{\Psi_q^k\big\}~,
 \label{enprod}
\end{equation}
with 

\begin{equation}
 \Psi_q^k=ln_q f_q^k-\bigg\{\frac{1\pm f_q^k}{f_q^k}\bigg\}^{q-1} ln_q(1\pm f_q^k)~.
 \label{shi}
\end{equation}

Substituting Eq.(\ref{releq}) into (\ref{enprod}), we achieve the entropy production in terms of collision term,

\begin{eqnarray}
 \sigma&&=-\sum_{k,l=1}^{N}\int\frac{d^3p_k}{(2\pi)^3p_k^0} C^{kl}_{q}\big\{\Psi^{k}_q\big\}\nonumber\\
  &&=-\frac{1}{2}\sum_{k,l,i,j=1}^{N}\int d\Gamma_k d\Gamma_l d\Gamma_i d\Gamma_j \big\{ \Psi^{k}_q \big\}\nonumber\\
  &&\times \bigg\{h_q\big[f_q^i,f_{q}^{j},(1\pm f_q^{k}),(1\pm f_{q}^{l})\big]W_{i,j|k,l}\nonumber\\
&&-h_q\big[f_q^k,f_{q}^l,(1\pm f_q^i),(1\pm f_{q}^j)\big]W_{k,l|i,j}\bigg\},
  \label{enprod1}
\end{eqnarray}
with the phase space factors now on denoted by $d\Gamma_{i}=\frac{d^3p_i}{(2\pi)^3p_i^0}$.
Interchanging the initial and final integration variables in the last term of eq.(\ref{enprod1}) and noting the transition rate to be symmetric
under the exchange of index pair $(k,l)$ and $(i,j)$, the expression for entropy production finally reduces to,

\begin{eqnarray}
 \sigma=&&-\frac{1}{4} \sum_{k,l,i,j=1}^{N}\int d\Gamma_k d\Gamma_l d\Gamma_i d\Gamma_j\nonumber\\
        &&\times\big\{\Psi^{k}_{q}+\Psi^{l}_{q}-\Psi^{i}_{q}-\Psi^{j}_{q}\big\}\nonumber\\
        &&\times \bigg\{h_q\big[f_q^i,f_{q}^{j},(1\pm f_q^{k}),(1\pm f_{q}^{l})\big]W_{i,j|k,l}\bigg\}~.
\label{sigma1}        
\end{eqnarray}

From the bilateral normalization property of the transition rate we have,

\begin{equation}
 \sum_{i,j=1}^{N}\int d\Gamma_i d\Gamma_i W_{k,l|i,j}=\sum_{i,j=1}^{N}\int d\Gamma_i d\Gamma_i W_{i,j|k,l}~.
 \label{sigma2}
\end{equation}
Multiplied both side of Eq.(\ref{sigma2}) by $h_q\big[f_q^k,f_{q}^{l},(1\pm f_q^{i}),(1\pm f_{q}^{j})\big]$ and integrating over $d\Gamma_{k}$
and $d\Gamma_{l}$ and then interchanging $(k,i)$ and $(l,j)$ in the left hand side after summing over $k$ and $l$, we end with the following relation,
\begin{eqnarray}
 &&\sum_{k,l,i,j=1}^{N}\int d\Gamma_k d\Gamma_l d\Gamma_i d\Gamma_j\bigg\{
 \frac{h_q\big[f_q^k,f_{q}^l,(1\pm f_q^i),(1\pm f_{q}^j)\big]}{h_q\big[f_q^i,f_{q}^j,(1\pm f_q^k),(1\pm f_{q}^l)\big]}\nonumber\\
 &&~~~~~~-1\bigg\} h_q\big[f_q^i,f_{q}^{j},(1\pm f_q^{k}),(1\pm f_{q}^{l})\big]W_{i,j|k,l}=0~.
 \label{sigma3}
\end{eqnarray}
Multiplying Eq.(\ref{sigma3}) with $1/4$ and adding it to the right hand side of Eq.(\ref{sigma1}), we are finally left with the final expression for entropy production,

\begin{eqnarray}
 \sigma=&&\frac{1}{4}\sum_{i,j,k,l=1}^{N}\int d\Gamma_k d\Gamma_l d\Gamma_i d\Gamma_j\big[A-ln_qA-1\big]\nonumber\\
 &&\times h_q\big[f_q^i,f_{q}^{j},(1\pm f_q^{k}),(1\pm f_{q}^{l})\big]W_{i,j|k,l}~,
\end{eqnarray}
with,
\begin{eqnarray}
 A=\frac{h_q\big[f_q^k,f_{q}^l,(1\pm f_q^i),(1\pm f_{q}^j)\big]}{h_q\big[f_q^i,f_{q}^j,(1\pm f_q^k),(1\pm f_{q}^l)\big]}\nonumber\\
 =exp_{q}\big[\Psi^{k}_{q}+\Psi^{l}_{q}-\Psi^{i}_{q}-\Psi^{j}_{q}\big]~.
\end{eqnarray}
The function $A-ln_qA-1$ is always positive for positive $A$ and positive $q$. It vanishes if and only if $A$ is
equal to 1, i.e, production rate of q-generalized entropy is always positive.
So at equilibrium $\sigma=0$, following $A=1$ we obtain $ln_qA=0$, i.e, $\Psi^{k}_{q}+\Psi^{l}_{q}=\Psi^{i}_{q}+\Psi^{j}_{q}$.
This clearly reveals $\Psi_q$ is a summation invariant. Now following the basic definition of summation invariant $\Psi_q$
is constructed as a linear combination of a constant and the 4-momenta $p_{k}^{\mu}$. This condition provides the structural
equation defining space-time dependence of distribution function in the following way,

\begin{equation}
ln_q f_q^k-\bigg\{\frac{1\pm f_q^k}{f_q^k}\bigg\}^{q-1} ln_q(1\pm f_q^k)=a_{k}(x)+b_{\mu}(x)p^{\mu}_{k}~. 
\label{dist2}
\end{equation}
Here $a_{k}$ and $b^{\mu}$ are space-time dependent arbitrary quantities except the constraint that the function $a_{k}(x)$
must be additively conserved, i.e, $a_{k}(x)+a_{l}(x)=a_{i}(x)+a_{j}(x)$. The 4-momentum conservation is always implied $p_{k}^{\mu}+p_{l}^{\mu}=p_{i}^{\mu}+p_{j}^{\mu}$.
Eq.(\ref{dist2}) after a few steps of simplification gives the expression for a q-generalized BE-FD single
particle distribution function belonging to $k^{th}$ species for a multicomponent relativistic system by the following equation,

\begin{equation}
f^{k}_{q}=\frac{1}{\big[1+(q-1)\big\{\frac{p_{k}^{\mu}u_{q\mu}}{T_q}-\frac{\mu_{kq}}{T_q}\big\}\big]^{\frac{1}{q-1}}\mp1}~,
\label{dist3}
\end{equation}
where we can make the identification 
\begin{equation}
 a_{k}=\frac{\mu_{kq}}{T_q},~~~~~~~~~~b^{\mu}=-\frac{u_{q}^{\mu}}{T_q}~,
\end{equation}
with $\mu_{kq}$ is the chemical potential for each particle of the $k^{th}$ species, $T_{q}(x,q)$ is the bulk temperature and 
$u_{q}^{\mu}(x,q)$ is the hydrodynamic 4-velocity of the fluid system under non-extensive environment. Hence $T_q$, $u_q^{\mu}$ and $\mu_{kq}$ 
are the intrinsic parameters of a q-equilibrated thermodynamic system. One can readily notice Eq.(\ref{dist3}) invariably leads to
Eq.(\ref{dist1}) with particle species $k$ for a co-moving frame of the fluid ($u_{q}^{\mu}(x)=(1,0,0,0)$).
This identification provides the necessary conformation about the form of entropy and collision integral
defined within the scope of covariant kinetic theory.

On the foundation of the formalism developed so far it is the time to define the basic macroscopic variables in the language
of relativistic kinetic theory. For this we start with the transport equation (\ref{releq}) itself. First Eq.(\ref{releq}) is
integrated over $\frac{d^3p_{k}}{(2\pi)^3p_k^0}$ and then summed over $k=0,N$. By the virtue of summation invariance the 
zeroth moment of collision term vanishes on the right hand side while the left hand side gives the conservation law as,
\begin{equation}
 \partial_{\mu}N^{\mu}_q(x,q)=0~,
 \label{conpn}
\end{equation}
where
\begin{equation}
 N_q^{\mu}=\sum_{k=1}^{N}g_k\int\frac{d^3p_{k}}{(2\pi)^3p_k^0}p_{k}^{\mu}\big\{f_{q}^{k}(x,p_{k},q)\big\}^q~,
 \label{pff}
\end{equation}
is defined as the q-generalized total particle 4-flow with $g_k$ as the degeneracy number.
The particle 4-flow $N_q^{\mu}$ must be proportional to the hydrodynamic 4-velocity $u_{q}^{\mu}$ since the equilibrium distribution
function $f_{q}^{k}$ singles out this particular direction in space-time. Thus the macroscopic definition of particle 4-flow
is given by,
\begin{equation}
 N^{\mu}_{q}(x,q)=n_{q}u_{q}^{\mu}~.
 \label{pffm}
\end{equation}
The proportionality factor is the q-generalized particle number density which can be expressed as,
\begin{eqnarray}
 n_q(x,q)=&&N_q^{\mu}u_{q\mu}\nonumber\\
 =&& \sum_{k=1}^{N}g_k\int\frac{d^3p_{k}}{(2\pi)^3p_{k}^{0}}p^{\mu}_{k}u_{q\mu}\big(f_{q}^{k}\big)^{q}~.
 \label{pn}
\end{eqnarray}
Applying the conservation law (\ref{conpn}) to the expression (\ref{pffm}) we finally obtain the equation of continuity
for a q-equilibrated system as the follows,
\begin{equation}
Dn_{q}=-n_{q}\partial_{\mu}u_{q}^{\mu}~, 
\end{equation}
with $D=u^{\mu}_{q}\partial_{\mu}$ as the covariant time derivative in the local rest frame. 

Next, same interaction and summation is performed over Eq.(\ref{releq}) but after multiplying with particle 4-momentum $p_{k}^{\mu}$
which again reduces the right hand side to zero since the first moment of collision term vanishes as well under the principle of
summation invariance. The resulting equation gives the energy-momentum conservation law in the following manner,
\begin{equation}
\partial_{\mu}T_{q}^{\mu\nu}(x,q)=0~,
\label{conem}
\end{equation}
where
\begin{equation}
 T_q^{\mu\nu}=\sum_{k=1}^{N}g_k\int\frac{d^3p_{k}}{(2\pi)^3p_k^0}p_{k}^{\mu}p_{k}^{\nu}\big\{f_{q}^{k}(x,p_{k},q)\big\}^q~,
\label{emf}
 \end{equation}
is defined as the q-generalized energy-momentum tensor. Noting it as a rank-2 tensor, its macroscopic definition is expressed
in terms of the available rank-2 tensors at our exposal, i.e, $u_{q}^{\mu}u_{q}^{\nu}$ and $g^{\mu\nu}$ in the following
way,

\begin{equation}
 T_q^{\mu\nu}=\epsilon_q u_{q}^{\mu}u_{q}^{\nu}-P_{q}\Delta_{q}^{\mu\nu}~,
 \label{emfm}
\end{equation}
with $\Delta_{q}^{\mu\nu}=g^{\mu\nu}-u_{q}^{\mu}u_{q}^{\nu}$ as the projection operator. The metric $g^{\mu\nu}$ of the system
is defined here as, $g^{\mu\nu}=(1,-1,-1,-1)$.
The q-generalized energy density $\epsilon_q$ and pressure $P_{q}$ is defined as,

\begin{eqnarray}
\epsilon_{q}=&&T_{q}^{\mu\nu}u_{q\mu}u_{q\nu}\nonumber\\
=&&\sum_{k=1}^{N}g_k\int\frac{d^3p_{k}}{(2\pi)^3p_{k}^{0}} \big\{p^{\mu}_{k}u_{q\mu}\big\}^2\big(f_{q}^{k}\big)^{q}~,
\label{ed1}\\
P_{q}=&&-\frac{1}{3}T_q^{\mu\nu}\Delta_{q\mu\nu}\nonumber\\
=&&-\frac{1}{3}\sum_{k=1}^{N}g_k\int\frac{d^3p_{k}}{(2\pi)^3p_{k}^{0}} p_{k}^{\mu}p_{k}^{\nu} \Delta_{q\mu\nu} \big(f_{q}^{k}\big)^{q}~.
\label{P1}
\end{eqnarray}
Applying the conservation law (\ref{conem}) to the expression (\ref{emfm}) and then contracting with $u_{q}^{\mu}$ and $\Delta_{q}^{\mu\nu}$
from left we finally result into equation of energy and and equation of motion respectively as follows,
\begin{eqnarray}
 De_q=-\frac{P_q}{n_q}\partial_{\mu}u_{q}^{\mu}~,\\
 Du_{q}^{\mu}=\frac{1}{n_qh_q}\nabla_{\mu}P_{q}^{\mu}~,
\end{eqnarray}
where $e_{q}=\frac{\epsilon_q}{n_q}$ and $h_q=e_q+\frac{p_q}{n_q}$ are the total energy and total enthalpy per particle respectively. 
For a multi component system they are defined as $e_q=\sum_{k=1}^{N}x_{kq}e_{kq}$ and $h_q=\sum_{k=1}^{N}x_{kq}h_{kq}$, 
where $e_{kq}$ and $h_{kq}$ are the same for $k^{th}$ species along with $x_{kq}=\frac{n_{kq}}{n_q}$ as the particle fraction
for the $k^{th}$ species.
Finally putting the expression of q-distribution function from (\ref{dist3}) into the entropy expression (\ref{ent-1})
and contracting with $u_{q}^{\mu}$, we obtain definition of entropy density in terms of macroscopic thermodynamic
quantities defined so far,
\begin{eqnarray}
 S_{q}=&&S_{q}^{\mu}u_{q\mu}\nonumber\\
 =&&\frac{\epsilon_q+P_q}{T_q}-\frac{n_q \mu_q}{T_q}~,
 \label{ent_den}
\end{eqnarray}
with $\mu_q=\sum_{k=1}^{N}x_{kq}\mu_{kq}$ as the total chemical potential per particle of the system.
It is interesting that similar conservation laws and equations of motion have also been observed in \cite{Wilk6,Biro-Molner1} discussing  
the non-extensive hydrodynamics for relativistic systems in Boltzmann generalization.

\subsection{Thermodynamics quantities in q-generalized BE and FD system}

The very basic technique of determining the thermodynamic quantities given in Eq.(\ref{pn}), (\ref{ed1}) and (\ref{P1}),
lies in performing the moment integral $\int\frac{d^3p_k}{(2\pi)^3p_k^0}p_{k}^{\mu}\cdots \big\{f_{q}^{k}\big\}^q$
over the $q^{th}$ power of the non-extensive distribution function (\ref{dist3}). To obtain a convenient way to
execute that, here an useful identity is being provided. First Eq.(\ref{dist3}) can be simply expressed in the following way,
\begin{equation}
 f_{q}^{k}=\frac{1}{exp\big[\frac{1}{(q-1)}ln\{1+(q-1)y_k\}\big]\mp1}~,
 \label{dist4}
\end{equation}
denoting $y_k=\frac{p_{k}\cdot u_q-\mu_{kq}}{T_q}$.
Eq.(\ref{dist4}) readily leads to its derivative in the following form,
\begin{equation}
\frac{\partial f^{k}_{q}}{\partial y_k}=-f^{k}_{q}\big(1\pm f^{k}_{q}\big)\frac{1}{\{1+(q-1)y_k\}}~.
\label{iden1}
\end{equation}
Again from (\ref{dist4}), the argument can be extracted in the following way,
\begin{equation}
 \bigg\{\frac{f_q^k}{1\pm f_q^k}\bigg\}^{(1-q)}=\big\{1+(q-1)y_k\big\}~.
 \label{iden2}
\end{equation}
Comparing Eq.(\ref{iden1}) and (\ref{iden2}), we obtain the following identity,
\begin{equation}
 \big\{f_{q}^{k}\big\}^q=\mp\frac{\partial}{\partial y_k}\bigg\{ \frac{(1\pm f_q^k)^{q-1}}{(q-1)} \bigg\}~.
 \label{iden3}
\end{equation}
Now the $\big(1\pm f_q^k\big)^{q-1}$ term can be expressed in an infinite series for small $(q-1)$ values,
(such that quadratic terms $\sim (q-1)^2$ are being neglected) in the following manner,

\begin{equation}
 \big(1\pm f_q^k\big)^{q-1}=1+(q-1)\sum_{l=1}^{\infty}(\pm)^l \frac{\big\{F_k exp(-y_k)\big\}^l}{l}~,
 \label{iden4}
\end{equation}
with $F_k=1-\frac{1}{2}(1-q)y_k^2$.
Upon taking derivative of Eq.(\ref{iden4}) and with the virtue of identity (\ref{iden3}), for a constant value of $\mu_{kq}$
we obtain $\big\{f^k_q\big\}^q$ in an infinite series as the following,
\begin{eqnarray}
 &&\big\{f^k_q\big\}^q=\sum_{l=1}^{\infty}(\pm)^{l-1}e^{l\tilde\mu_k}e^{-l\tau_k}+(q-1)\nonumber\\
 &&\times\sum_{l=1}^{\infty}(\pm)^{l-1}e^{l\tilde\mu_k}e^{-l\tau_k}
\bigg\{\frac{l}{2}\tilde{\mu_{k}}^2+\tilde{\mu_{k}}-l\tilde{\mu_{k}}\tau_k-\tau_k+\frac{l}{2}\tau_{k}^2\bigg\}.\nonumber
\\
\label{dist5}
\end{eqnarray}
Here, $\tau_k=\frac{p_k\cdot u_q}{T_q}$. From now on wards we will denote $T_q$ and $\mu_{kq}$ simply by $T$ and $\mu_k$ for 
convenience. Details of the derivations
are given in Appendix. For the mathematical properties of q-logarithm and q-exponential functions Ref.\cite{Yamano} has been essentially
helpful. So one can see Eq.(\ref{dist5})
contains a series term contributing in the usual BE-FD integrals and another series term proportional to $(q-1)$ that
contributes in the non-extensivity while determining the thermodynamic quantities. In the Boltzmann limit Eq.(\ref{iden3})
simply becomes,

\begin{equation}
 \big\{f_{q}^{k}\big\}_{B}^q=-\frac{\partial}{\partial y_k}\big\{f_{q}^{k}\big\}_{B}~,
 \label{iden5}
\end{equation}
which finally can be expressed as,
\begin{eqnarray}
 \big\{f^k_q\big\}_{B}^q=&&e^{\tilde\mu_k}e^{-\tau_k}\nonumber\\
 +&&(q-1)\bigg[ \frac{1}{2}\tilde{\mu_{k}}^2+\tilde{\mu_{k}}-\tilde{\mu_{k}}\tau_k-\tau_k+\frac{1}{2}\tau_{k}^2 \bigg],
 \label{dist6}
\end{eqnarray}
with the q-generalized Boltzmann distribution function as
\begin{eqnarray}
 \big\{f^k_q\big\}_{B}=&&exp_{q}(-(\tau_k-\tilde{\mu}_k))~,\nonumber\\
                      =&&\big\{1+(q-1)(\tau_k - \tilde \mu_k)\big\}~.
 \label{Boltz_dist}                     
\end{eqnarray}

Now one can readily note that Eq.(\ref{dist5}) reduces to Eq.(\ref{dist6}), if the series truncates at the leading term ($l=1$) only.
However the higher order terms resulting from the quantum correlations will be observed to have significant effect (at least at quantitative
level) on the thermodynamic variables in the next section.

Hence putting the form of $\big\{f_{q}^{k}\big\}^q$ from Eq.(\ref{dist5}) into the momentum integral (\ref{pn}),(\ref{ed1}) and (\ref{P1}),
the expressions for particle number density, energy density and pressure can be achieved. While for massive hadron gas the integrals reduced
to infinite series over modified Bessel function of second kind, for a massless quark gluon gas it reduces to Polylog functions. The complete
analytical expressions for the two cases are given below.

\subsubsection{Massive hadron gas with non-zero baryon chemical potential}
 For this case first the infinite series has been defined for the BE and FD case respectively as the following,
 \begin{equation}
  S_{kn}^{\alpha}=\sum_{l=1}^{\infty}(\pm)^{l-1}exp(l\tilde{\mu}_{k})\frac{1}{l^{\alpha}}K_{n}(lz_{k}),
  \label{Poly}
 \end{equation}
with $z_{k}=\frac{m_k}{T_q}$, $m_k$ being the mass of $k^{th}$ hadron and $K_{n}(lz_{k})$ is the 
modified Bessel function of second kind with order $n$ and argument $lz_{k}$ defined as,
\begin{equation}
 k_{n}(z)=\frac{2^n n!}{(2n)!}\frac{1}{z^n}\int_{z}^{\infty}d\tau (\tau^2-z^2)^{n-\frac{1}{2}}exp(-\tau)~.
\end{equation}
The $\pm$ sign in Eq.(\ref{Poly}) respectively indicates bosonic and fermionic hadrons.
Following these notations here, the analytical results for macroscopic thermodynamic quantities of a massive hadron gas 
in non-extensive environment have been given.

Expression for particle number density :
\begin{eqnarray}
 \frac{n_q}{\{\frac{T^3}{2\pi^2}\}}&&=\sum_{k=1}^{\infty} g_k \bigg[\big(z_k^2 S_{k2}^{1}\big)\nonumber\\
  &&+(q-1)\bigg\{ \tilde{\mu}_k^2 \big(\frac{1}{2}z_{k}^{2}S_{k2}^{0}\big)-\tilde{\mu}_{k}\big(2z_k^2S_{k2}^{1}+z_{k}^{3}S_{k1}^{0}\big)\nonumber\\
  &&+\big(3z_k^2S_{k2}^{2}+\frac{3}{2}z_{k}^{3}S_{k1}^{1}+\frac{1}{2}z_{k}^{4}S_{k0}^{0}\big)\bigg\}\bigg]
  \label{pnh}
\end{eqnarray}

Expression for energy density :
\begin{eqnarray}
\frac{\epsilon_q}{\{\frac{T^4}{2\pi^2}\}}&&= \sum_{k=1}^{\infty} g_k \bigg[\big(3z_k^2 S_{k2}^2+z_k^3S_{k1}^{1}\big)\nonumber\\
&&+(q-1)\bigg\{ \tilde{\mu}_k^2 \big( \frac{3}{2}z_k^2 S_{k2}^1+\frac{1}{2}z_k^3 S_{k1}^{0}\big)\nonumber\\
&& +\tilde{\mu}_k \big( -9z_k^2S_{k2}^2-4z_k^3S_{k1}^{1}-z_k^4S_{k0}^0 \big)\nonumber\\
&& +\big(18z_k^2S_{k2}^{3}+\frac{15}{2}z_{k}^3S_{k1}^2+\frac{3}{2}z_k^4S_{k0}^{1}\nonumber\\
&& +\frac{1}{2}z_k^4 S_{k2}^1+\frac{1}{2}z_k^5 S_{k1}^{0}\big)\bigg\}\bigg]
\label{edh}
\end{eqnarray}

Expression for pressure :
\begin{eqnarray}
 \frac{P_q}{\{\frac{T^4}{2\pi^2}\}}&&=\sum_{k=1}^{\infty} g_k \bigg[\big(z_k^2 S_{k2}^{2}\big)\nonumber\\
  &&+(q-1)\bigg\{ \tilde{\mu}_k^2 \big(\frac{1}{2}z_{k}^{2}S_{k2}^{1}\big)-\tilde{\mu}_{k}\big(3z_k^2S_{k2}^{2}+z_{k}^{3}S_{k1}^{1}\big)\nonumber\\
  &&+\big(6z_k^2S_{k2}^{3}+\frac{5}{2}z_{k}^{3}S_{k1}^{2}+\frac{1}{2}z_{k}^{4}S_{k0}^{1}\big)\bigg\}\bigg]
  \label{Ph}
\end{eqnarray}

\subsubsection{Massless QGP with non-zero quark chemical potential}

For this case first the PolyLog function is defined as the following,
\begin{equation}
\sum_{k=1}^{\infty} (\pm)^{k-1}\frac{e^{\pm k\tilde{\mu}}}{k^a}=\pm\textrm{PolyLog}[a,\pm e^{\pm\tilde{\mu}}]~.
\label{Polylog1}
\end{equation}
Further for small quark chemical potential $\mu$, keeping only upto terms of the order of $\tilde{\mu}^2$ the Polylogs
satisfy the following identity,
\begin{eqnarray}
 \textrm{PolyLog}[a,-e^{\pm\tilde{\mu}}]&&=\textrm{PolyLog}[a,-1]\nonumber\\
 &&\pm\tilde{\mu}~\textrm{PolyLog}[(a-1),-1]\nonumber\\
 &&+\frac{\tilde{\mu}^2}{2}\textrm{PolyLog}[(a-2),-1].
 \label{Polylog2}
\end{eqnarray}
With the help of Eq.(\ref{Polylog1}) and (\ref{Polylog2}), for a system with massless quarks, antiquarks and gluons
the macroscopic thermodynamic quantities for a non-extensive environment are given in this section.

Expression for particle number density :
\begin{eqnarray}
\frac{n_q}{\{\frac{T^3}{2\pi^2}\}}&&=2\bigg[ g_g \zeta(3) -g_q \big\{2\textrm{PolyLog[3,-1]}-\tilde{\mu}^2 ln2\big\}\bigg]\nonumber\\
&&+(q-1)\bigg[ 6\big\{ g_g\zeta(4)-2g_q\textrm{PolyLog[4,-1]}\big\}\nonumber\\
&&+8\tilde{\mu}g_q\textrm{PolyLog[3,-1]}-8\tilde{\mu}^2g_q\textrm{PolyLog[2,-1]}\bigg] .\nonumber\\
\label{pnqg}
\end{eqnarray}

Expression for energy density and pressure:
\begin{eqnarray}
&&\frac{\epsilon_q}{\{\frac{T^4}{2\pi^2}\}}=\frac{3P_q}{\{\frac{T^4}{2\pi^2}\}}=\nonumber\\
&&6\bigg[ g_g \zeta(4) -g_q \big\{2\textrm{PolyLog[4,-1]} +\tilde{\mu}^2 \textrm{PolyLog[2,-1]}\big\}\bigg]\nonumber\\
&&+(q-1)\bigg[ 36\big\{ g_g\zeta(5)-2g_q\textrm{PolyLog[5,-1]}\big\}\nonumber\\
&&+36\tilde{\mu}g_q\textrm{PolyLog[4,-1]}-42\tilde{\mu}^2g_q\textrm{PolyLog[3,-1]}\bigg] ~.
\label{edqg}
\end{eqnarray}

$\zeta(n)$ are the Riemann zeta functions defined as $\zeta(n)=\textrm{PolyLog}[n,1]=\sum_{k=1}^{\infty}\frac{1}{k^n}$, and
$\textrm{PolyLog}[1,-1]=-ln2$. The gluon and quark/antiquark degeneracies are respectively given by $g_g=16$ and $g_q=2N_cN_f$,
where $N_c=3$ is the color and $N_f=2$ is the flavor degrees of freedom for the quarks.

\subsection{Effective coupling in q generalized hot QCD system}
After defining all the required thermodynamic state variables in a non-extensive environment, it is instructive to look for the Debye
mass and the effective coupling of the system under the same. Previously the Debye shielding has been discussed with the non-extensive
effects for an electron-ion plasma in \cite{Bouzit,Gougam}. Here the effects of non-extensivity is being observed on the Debye mass and
the effective coupling for an interacting quark-gluon plasma system.

Following the definition of Debye mass from semiclassical transport theory \cite{Litim,Kelly,Blaizot} the same has been defined for a 
non-extensive system as the following,

\begin{eqnarray}
 m_{D}^2|_q=-4\pi\alpha_s(T,\mu)\sum_{k=1}^{N}2C_k\int \frac{d^3{p_k}}{(2\pi)^3} \frac{\partial }{\partial |\vec{p_k}|}\{f_q^k\}^q~,
 \label{dm1}
\end{eqnarray}
with $C_k=N_c$ for gluons and $C_k=N_f$ for quark/antiquarks. Applying Eq. (\ref{dist5}) into (\ref{dm1}), we obtain the q-generalized
Debye mass value by the following expression,
\begin{eqnarray}
 m_D^2|_q =4\pi\alpha_s \frac{T^2}{\pi^2}\bigg[ 2N_c\zeta(2)-2N_f\textrm{PolyLog}[2,-1]+\frac{N_f}{2}\tilde{\mu}^2\nonumber\\
 +(q-1)\bigg\{ 2N_c \zeta(3)-2N_f\textrm{PolyLog}[3,-1]\bigg\}\bigg].\nonumber\\
 \label{dm2}
\end{eqnarray}
The first part of Eq. (\ref{dm2}) is simply the leading order HTL estimation of Debye mass,
\begin{equation}
 m_D^2|_{q\rightarrow1}=4\pi\alpha_s T^2 \bigg[ \bigg\{ \frac{N_c}{3}+\frac{N_f}{6}\bigg\}+\tilde{\mu}^2\frac{N_f}{2\pi^2}\bigg]~.
 \label{dm3}
\end{equation}
From Eq. (\ref{dm2}) and (\ref{dm3}), it is instructive to obtain the q-generalized effective coupling for a non-extensive system by the following
expression
 \begin{eqnarray}
  &&\alpha_q(T,\mu,q)=\alpha_s(T,\mu)\nonumber\\
  &&\times\bigg[1+(q-1)\frac{\big\{\frac{2N_c}{\pi^2} \zeta(3)-\frac{2N_f}{\pi^2} \textrm{PolyLog}[3,-1]\big\}}
  {\big\{\frac{N_c}{3}+\frac{N_f}{6}\big\}+\tilde{\mu}^2 \big\{\frac{N_f}{2\pi^2}\big\}}\bigg].
  \label{alpha_eff}
 \end{eqnarray}
Here $\alpha_s(T,\mu)$ is the QCD running coupling constant at finite temperature and quark chemical
potential. Here it's value has been set from the 2-loop QCD gauge coupling calculation at finite
temperature from the Ref.(\cite{Laine}).
\\
\section{Results and Discussions}

In this section, we proceed to analyze the quantitative impact of non-extensivity in small $(q-1)$ limit along
with the quantum corrections in terms of BE/FD statistics on various thermodynamic quantities, compared to the same
under Boltzmann generalization.The results for massive and massless systems are presented in two separate subsections.

Before proceeding for providing the results, it is essential to specify the value of the entropic parameter $q$
for systems likely to be created in relativistic heavy ion collisions. We must stick to the small ($q-1$) limit
in order to neglect the quadratic terms $\sim (q-1)^2$. In a number of works the value of $q$
has been attempted to obtain from the fluctuation of the system parameters like temperature or number concentration
\cite{Biro4,Biro5,Wilk4,Wilk2,Wilk1}. In \cite{Biro5} this value has been reported by a range $1.0<q<1.5$ for 
high-energy nuclear collisions. In \cite{Cleymans5}, an upper limit for $q$ is given by $q<\frac{4}{3}$ in order
to get a non-divergent particle momentum distribution $E\frac{dN}{d^3p}$. Apart of that in a number of 
literature the $q$ parameter has been extracted by fitting the transverse momentum spectra of final state
hadrons with the experimental data from ALICE, ATLAS and CMS \cite{Cleymans1,Cleymans2,Cleymans3,Cleymans4}.
Theses fitted values of $q$ parameter ranges from $1.110\pm0.218$ to $1.158\pm0.142$. In light of the above
observations, the $q$ parameter have been set at $q=1.15$ and $q=1.3$ as an intermediate and an extremal value,
apart of $q=1$ which corresponds to the ideal BE/FD case. 

\subsection*{Results of massive pion-nucleon system with non-zero baryon chemical potential}

\begin{figure}[h]
\includegraphics[scale=0.35]{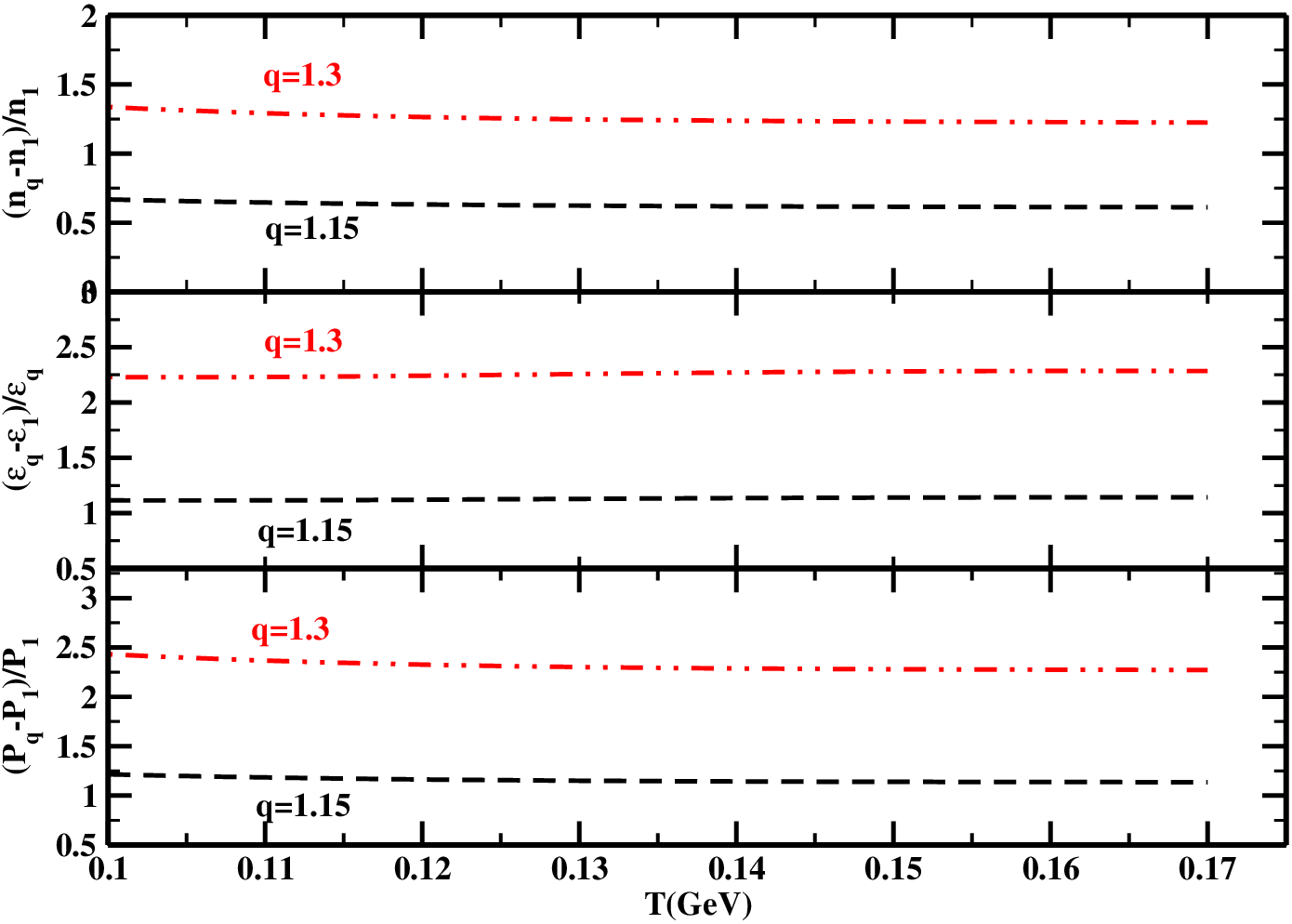}
\caption{Relative ratios of the non-extensive correction to the $q=1$ values of thermodynamic quantities.}
\label{Massive_full_rel}
\end{figure}

In this section, first the effects of non-extensivity on the macroscopic thermodynamic quantities have been shown
for a massive pion-nucleon gas with chemical potential for nucleons $\mu_N=0.1$ GeV. The temperature 
dependence of $(n_q-n_1)/n_1$, $(\epsilon_q-\epsilon_1)/\epsilon_1$ and $(P_q-P_1)/P_1$ have been plotted for 
$q=1.15$ and $q=1.3$ in 
three separate panels in Fig.(\ref{Massive_full_rel}) as obtained from Eq.(\ref{pnh}), Eq.(\ref{edh}) and 
Eq.(\ref{Ph}). Clearly $A_q (A\equiv n,\epsilon ,P)$ being the q generalized thermodynamic quantities 
(with quantum corrections) and $A_1$ being the same with $q=1$, i.e the ideal BE/FD quantities, the plotted
ratios express the relative significance of the non-extensive generalization of macroscopic quantities (with
BE/FD distributions) with respect to the ideal ones. From Fig.(\ref{Massive_full_rel}) the q-corrections
are appeared to be comparable with the quantities with ideal BE/FD distributions which further increases with
increasing $q$ values. These corrections are observed to be more pronounced for energy density and pressure than
the particle number density. Hence the effect of non-extensive generalization of single particle Bose-Einstein and Fermi-Dirac
distributions, is observed to produce significant effect while determining the thermodynamic quantities in a multi-component 
system, which is a massive pion-nucleon gas in the present case.
\\
\\

\begin{figure}[h]
\includegraphics[scale=0.35]{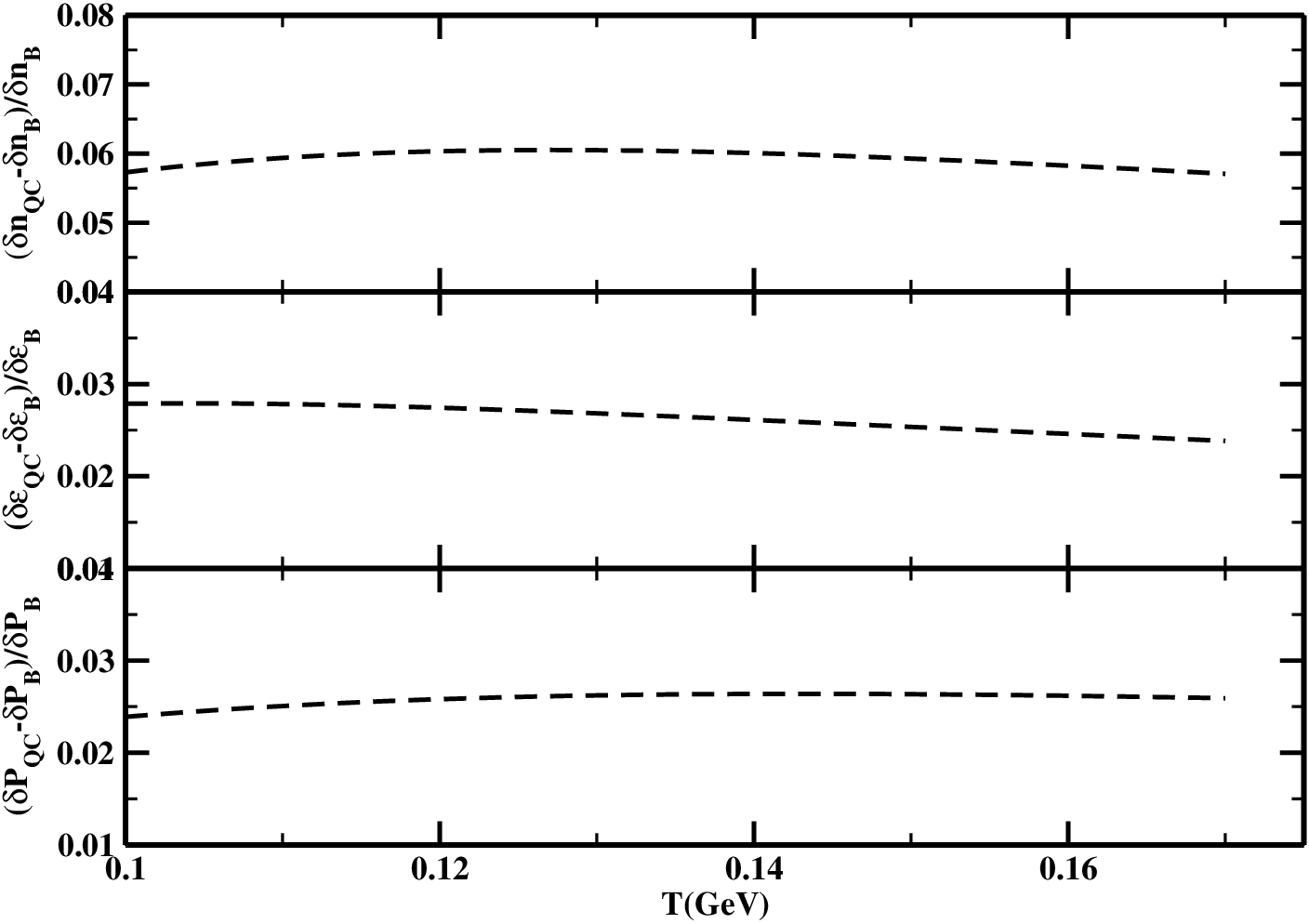}
\caption{Relative correction of non-extensive part using BE/FD distribution over Bolzmann statistics.}
\label{Massive_Rel_qterm}
\end{figure}

Next, in order to visualize the impact of non-extensivity on the quantum corrections, the nonextensive corrections ($A_q-A_1$) 
of particle number density, energy density and pressure obtained by 
generalizing the BE/FD distribution, relative to the same by generalizing Boltzmann distribution have
been given in Fig.(\ref{Massive_Rel_qterm}). $\delta A_{QC}$ denotes the terms proportional to ($q-1$) in
Eq.(\ref{pnh}), (\ref{edh}) and (\ref{Ph}) and $\delta A_{B}$ denotes the same with q-generalized Boltzmann
distribution given in Eq.(\ref{Boltz_dist}). The relative ratio shows the quantum correction taken in the 
nonextensive terms of thermodynamic quantities, makes how much difference compared to 
the ones without the quantum corrections. The relative change is $2-3\%$ for energy density and pressure 
which is displaying maximum impetus for particle number density upto $6\%$ for the massive $\pi-N$ system.  
\\
\\

\begin{figure}[h]
\includegraphics[scale=0.35]{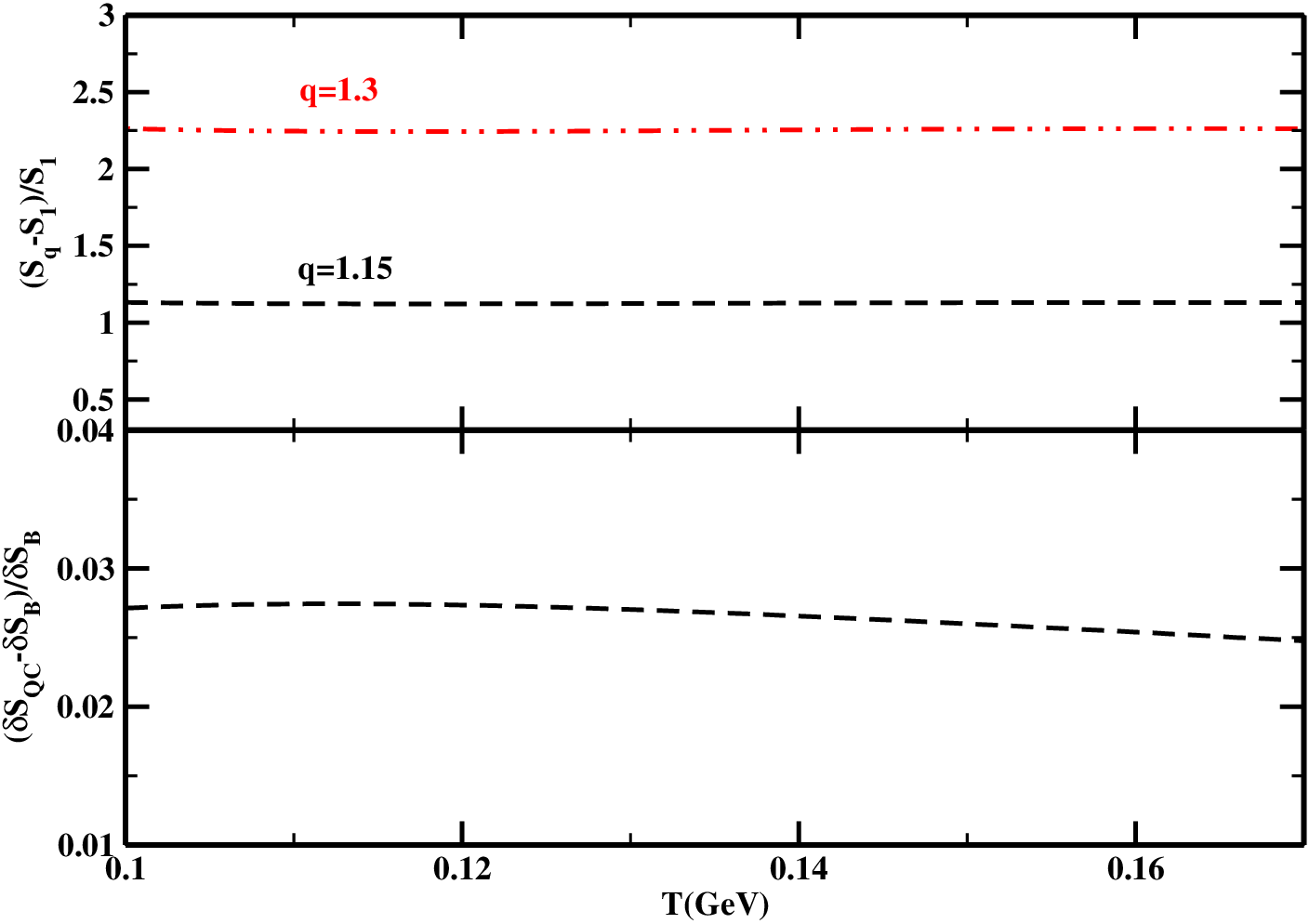}
\caption{The non-extensive correction and the quantum correction over non-extensivity of entropy of the system. }
\label{Massive_ent}
\end{figure}

Finally both the significance of the non-extensivity (for $q=1.15$ and $q=1.3$) and quantum corrections
of entropy density for the $\pi-N$ system following from Eq.(\ref{ent_den}) have been depicted in 
Fig.(\ref{Massive_ent}) in two separate panels. The effect of non-extensive terms in entropy appears 
to be of the same order of the leading term itself. The quantum correction in the non-extensive terms 
shows $\sim 3\%$ increment compare to the Bolzmann generalization.
\\
\\

\begin{figure}[h]
\includegraphics[scale=0.35]{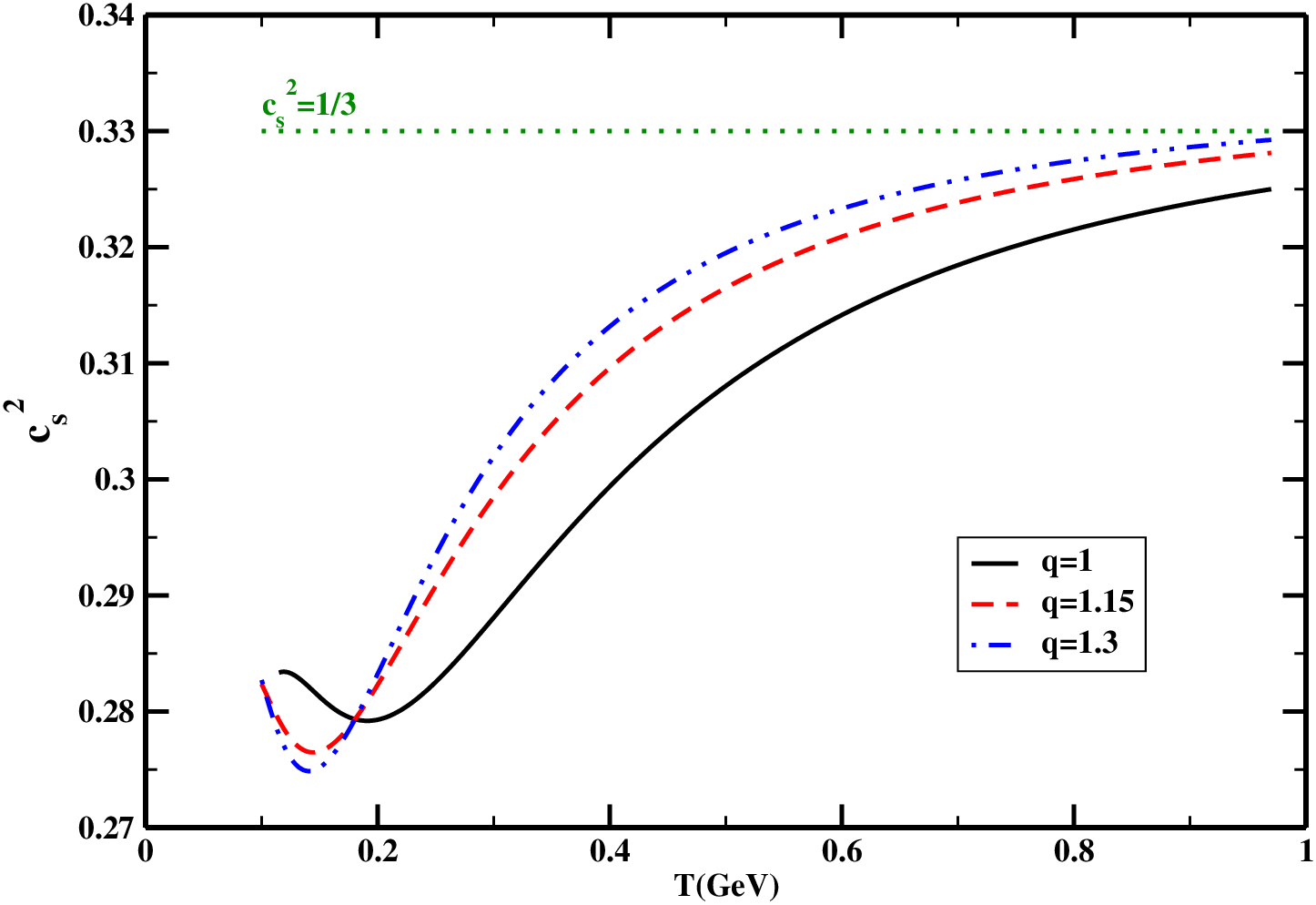}
\caption{Velocity of sound as a function of temperature in non-extensive environment. }
\label{cs2}
\end{figure}

After discussing the basic thermodynamic quantities, the square of sound velocity ($c_s^2$) has been plotted as a function of temperature for a 
massive pion-nucleon gas in a non-extensive environment along with the quantum corrections, for $q=1$, $q=1.15$ and $q=1.3$ in Fig.(\ref{cs2}).
For a system composed of massless particles the value of $c_s^2$ simply becomes $1/3$ (which is the Stefan-Boltzmann (SB) limit) irrespective of 
the value of $q$, which is depicted by the dotted straight line. For a massive pion-nucleon gas, $c_s^2$ shows the usual increasing trend with
increasing temperature for all values of $q$ which tends to approach the SB limit at considerably high temperatures. 
With higher $q$ values the speed of sound appears to become larger which is in accordance with Ref.\cite{Biro-Molner2} and approaches to the 
SB limit faster. This increment is much expected due to the significant increase in the system pressure in a non-extensive environment with 
respect to the ideal ($q=1$) one. The kink in the temperature dependence of $c_s^2$ (the minimum most point) below $0.2$ GeV agrees with the 
lattice results, where with increasing $q$ values the shift of the kink towards lower temperatures ($0.145-0.150$ GeV), indicates the fact that 
the minimum of the speed of sound, lies on the low temperature side of the crossover region \cite{Bazavov2014,Bazavov2009,Borsanyi2014,Borsanyi2012}.  
\\
\\

\subsection*{Results of massless quark-gluon system with non-zero quark chemical potential}

\begin{figure}[h]
\includegraphics[scale=0.35]{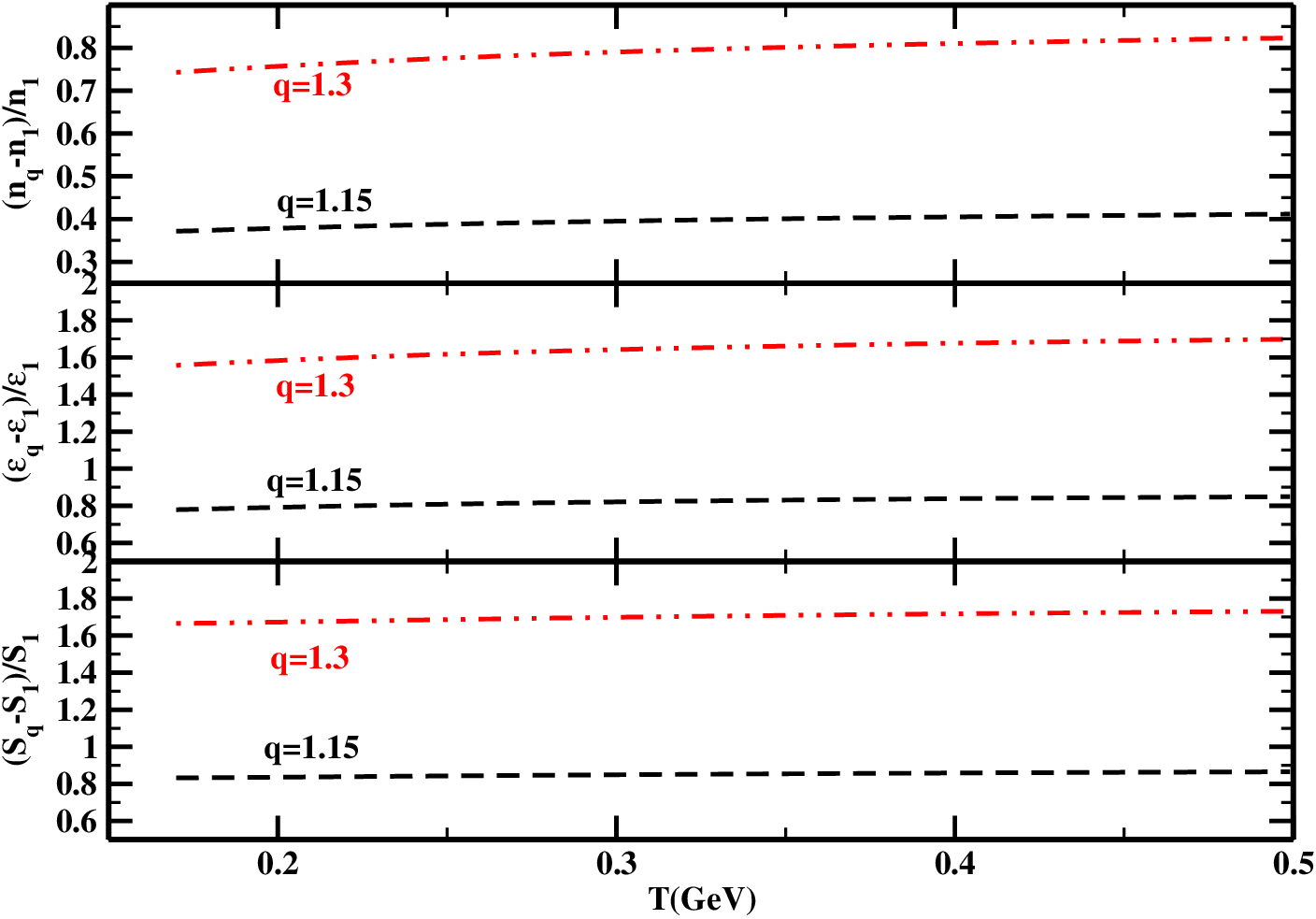}
\caption{Relative ratios of the non-extensive correction to the $q=1$ values of thermodynamic quantities.}
\label{Massless_full_rel}
\end{figure}

After presenting the results for the massive system, now we turn to the same for a massless quark-gluon system with
quark flavor $N_f=2$ and quark chemical potential $\mu=0.1$ GeV. Fig.(\ref{Massless_full_rel}) depicts the effect of 
q-parameters on the thermodynamic quantities (particle number density, energy density and
entropy as given in Eq.(\ref{pnqg}), Eq.(\ref{edqg}) and Eq.(\ref{ent_den})) estimated by generalizing BE/FD
distributions under a non-extensive environment. The $q$ corrections are comparable to the leading terms in this case
as well, however the increments are little less compared to that of the massive case. Clearly the effect of a finite
mass is introducing larger contributions to the terms proportional to $(q-1)$ for $q=1.15$ and $q=1.3$. However for 
a massless quark-gluon case the change in the thermodynamic quantities is quite significant as observed from
Fig.(\ref{Massless_full_rel}) and hence the effect of non-extensivity is proved to be quite relevant in the 
quark-gluon sector as well.
\\
\\

\begin{figure}[h]
\includegraphics[scale=0.35]{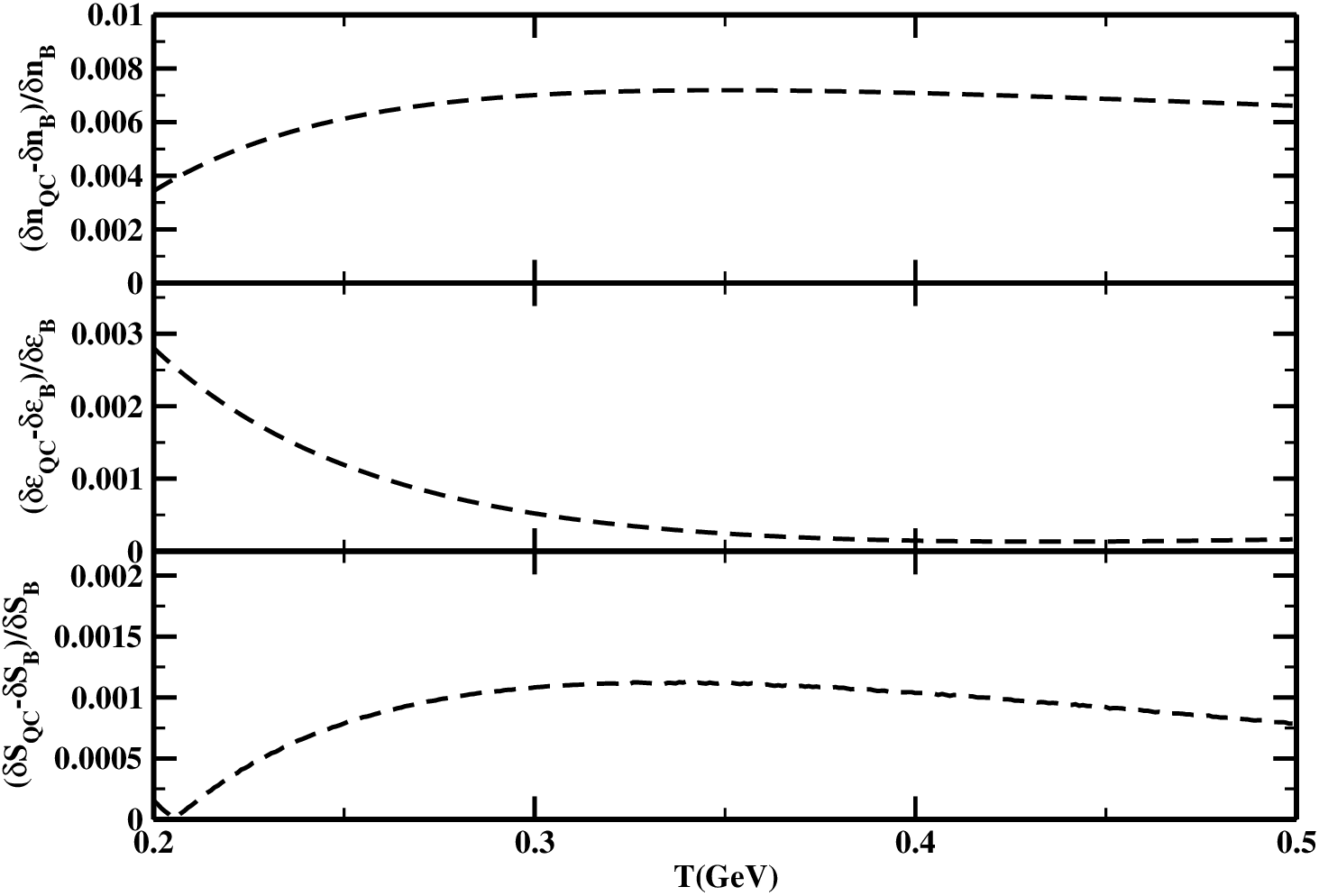}
\caption{Relative correction of non-extensive part using BE/FD distribution over Bolzmann statistics.}
\label{Massless_Rel_qterm}
\end{figure}

Secondly, the significance of the quantum corrections taken in the terms proportional to ($q-1$) (other than $q=1$)
while determining the thermodynamic quantities has been shown in Fig.(\ref{Massless_Rel_qterm}). The relative change
in non-extensive terms, taking the quantum corrections, as compared to the same by Bolzmann q-generalization is depicted 
by these ratios. However the quantum corrections in the high temperature, massless case is observed to be significantly
smaller than the massive case. This is quite anticipated because of the fact that at sufficiently high temperature
and low density, the quantum statistics reaches its classical limit and consequently the quantum distribution laws,
whether BE or FD, reduce to the Boltzmann distribution. Hence the reduction in the amount of change created by the 
quantum correction in the massless case where the temperature is well above $0.2$ GeV, is justified by the high
temperature behavior of BE/FD statistics where they scale down to the Bolzmann statistics. 
\\
 \\

\begin{figure}[h]
\includegraphics[scale=0.35]{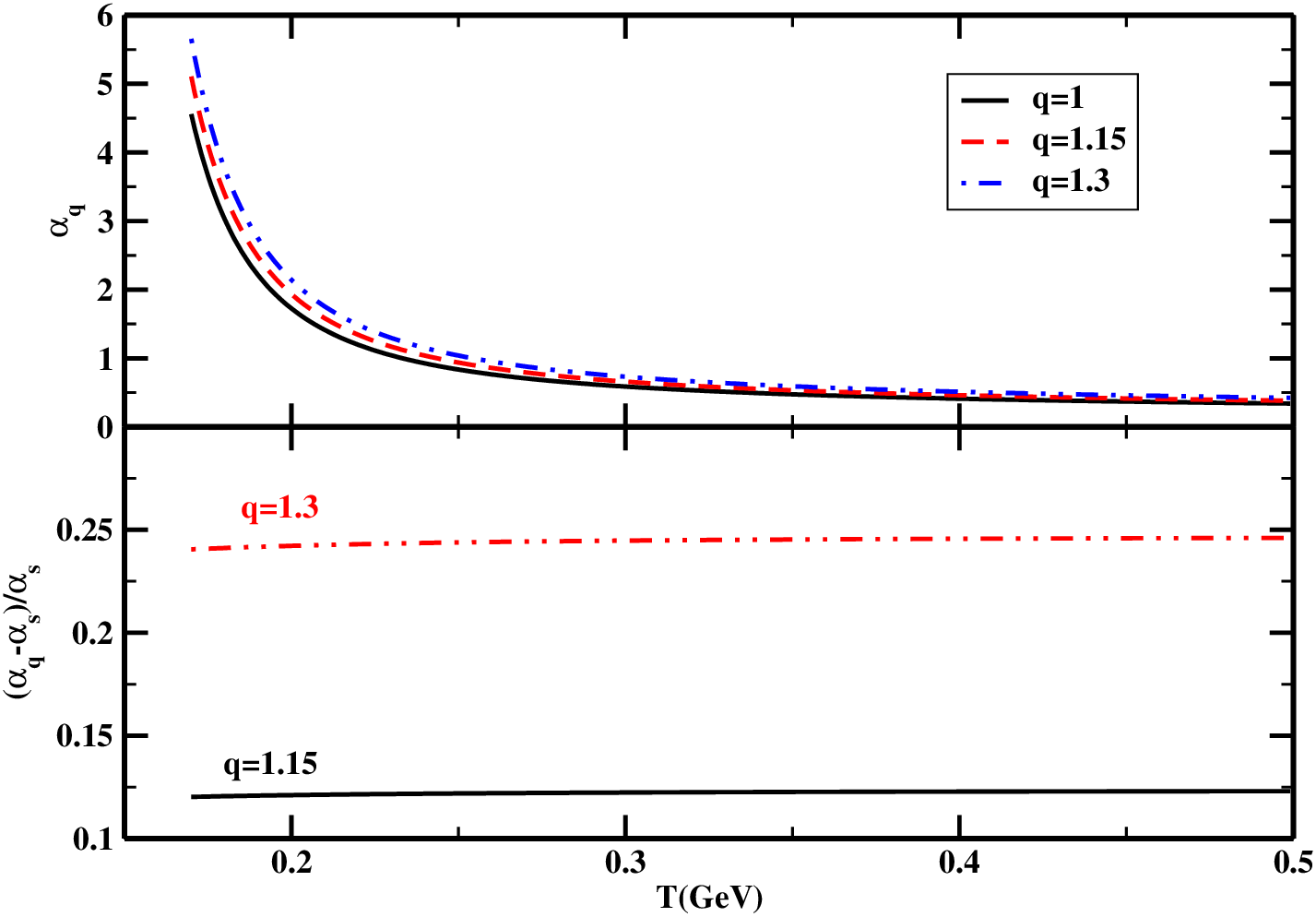}
\caption{The QCD coupling in a non-extensive environment as a function of temperature and its increment with respect to running coupling.}
\label{alpha}
\end{figure}

Finally, it is interesting to understand the impact of $q$-generalization on the hot QCD coupling in the quark-gluon
system and phenomenon such as Debye screening there. To that end, we proceed to investigate the QCD effective coupling
in a non-extensive QGP environment.
Following Eq.(\ref{alpha_eff}) the effective coupling $\alpha_q$ has been plotted as a function of temperature in the 
upper panel of Fig.(\ref{alpha}) for three values of $q$. The $q=1$ case simply gives the running coupling constant
$\alpha_s$,while $q=1.15$ and $q=1.3$ are showing the effects of non-extensivity which enhances the temperature
dependence of $\alpha_q$. This enhancement is mostly prominent around phase transition temperature ($0.17-0.2$ 
GeV) while in high temperature regions it tends to reduce its effects. In the lower panel to of Fig.(\ref{alpha}) the 
relative increment of $\alpha_q$ with respect to $\alpha_s$ has been shown, which is showing $7-25\%$ increment in 
the QCD coupling due to non-extensive effects. This significant increment is expected to improve the quantitative
estimates of the thermodynamic quantities which include the dynamical interactions and hence the coupling as the 
computational inputs, such as transport parameters of the system. Therefore we can conclude that, this dynamical modification
due to non-extensivity has far reaching effects on the viscosities and conductivities of the system and also on
their application in hydrodynamic simulations, in order to describe the space-time evolution of the system.
\\
\\

\section{Conclusion and Outlook}

In the present work, the relativistic kinetic theory and the thermodynamic properties have been obtained in detail by 
generalizing the Bose-Einstein and Fermi-Dirac distributions in an non-extensive environment following the prescription
introduced by Tsallis. Following the nonextensive definition of entropy provided by Tsallis, first, the q-generalized 
BE/FD distribution functions have been achieved from a grand canonical ensemble employing a number of constraints, namely,
norm constraint and constraints of internal energy and particle number, using the method of Lagrange's undetermined
multipliers. 

Next, the relativistic kinetic theory for a multi component system has been explored under the non-extensive
dynamics. Setting up the microscopic definition of q-entropy in terms of the single particle distribution function for a 
BE/FD system and defining a proper collision term under the same conditions, we finally achieve again the
expression for the q-generalized BE/FD distribution functions by using the techniques of entropy maximization and 
summation invariants. In a co-moving frame with the hydrodynamic velocity of the system, the obtained expression of the q-generalized 
BE/FD distribution function from relativistic kinetic theory reduces to the same obtained for grand canonical thermostatistics, proving the congruity
of the microscopic definitions of the entropy and collision term that have been employed here.

After setting up the theoretical framework, the macroscopic state variables such as particle number density, energy density, pressure,
entropy density and the velocity of sound, have been determined with these single particle distribution functions for a 
massive pion-nucleon and a massless quark-gluon system with non-zero
baryon chemical potential in small $(q-1)$ limit. Furthermore, the Debye mass and the effective coupling for an interacting QCD 
system have been estimated indicating the dynamical behavior of the system under the non-extensive generalization. 
The macroscopic thermodynamic quantities show significant increment due to the inclusion of non-extensive term for 
both the systems, which seems to be more dominant in the massive case. The relative change in the non-extensive terms due
the BE/FD generalization over Bolzmann distribution ranges from $2-6\%$ in hadronic system, which reduces to less than
$1\%$ for a quark-gluon system at higher temperature. For larger $q$ values $c_s^2$ enhances, subsequently reaching the SB limit
faster. Due to the non-extensive generalization, the temperature dependence of QCD coupling is observed to enhance 
significantly over the running coupling constant which is becomes ever larger for higher $q$ values.

The present work opens up a number of possible horizons to be explored under the non-extensive generalization of the 
system properties concerning heavy ion physics. One immediate future project to be investigated, is the transport
parameters of the system using the current formalism. Transport coefficients being crucial inputs in the hydrodynamic
equations describing system's space-time evolution, their response to a non-extensive medium where long range
correlations and memory effects are significant, is a highly interesting topic to venture in near future.
Anticipating the non-extensive formalism to provide a closer look to the systems created in heavy ion collisions,
setting up the hydro equations under this construction and their solution to obtain the space-time behavior of 
system temperature and hydrodynamic velocity are also extremely essential. A few studies in this regard has been done in
\cite{Biro-Molner2,Wilk6}. More extensive studies including the proper dynamics is an essential future task
in this line of work.

Finally, an extended study concerning the microscopic dynamics of the system under the application of non-extensivity
from a first principle approach is extremely necessary. In\cite{Biro6} an effective field theory has been discussed to
describe the nuclear and quark matter at high temperature by extending the Boltzmann-Gibbs canonical view
to Tsallis approach. A perturbation treatment of relativistic quantum field systems within the framework of Tsallis 
statistics have been studied in \cite{Kohyama}. Following this line of works a complete study of the relativistic
field theory under the non-extensive framework, in order to describe the dynamical properties of QGP at finite
temperature and baryon chemical potential is the next urgent thing to look for. In \cite{Wilk7} an effective
theory has been modelled describing the interplay between the non-extensivity and the QCD strong interaction dynamics 
in terms of a quasi particle model. So developing the complete generalization of the hot QCD medium including 
long range interactions is one of the most prospectful works to pursue in the upcoming days. Inspired by that,
such a generalization with proper equations of state encoding the finite temperature medium effects from
latest perturbative HTL calculations and lattice simulations are the immediate projects to be explored in near
future.

\appendix
\section{Details of the $\big\{f_q^k\big\}^q$ derivation}
Following the definition of $f_{q}^k$ from (\ref{dist3}) first we obtain,

\begin{eqnarray}
 (1\pm f_q^k)^{q-1}&&=\big[1\mp\big\{1+(q-1)y_k \big\}^{\frac{1}{1-q}}\big]^{1-q}\nonumber\\
                  &&=exp\bigg[(1-q)ln\big[1\mp X_k \big]\bigg]~,
 \label{app1}
 \end{eqnarray}
with,
\begin{eqnarray}
 X_k&&=\big[1+(q-1)y_k \big]^{\frac{1}{1-q}}\nonumber\\
 &&exp\big[\frac{1}{1-q}ln\{1+(q-1)y_k\}\big]~.
 \label{app2}
\end{eqnarray}
Expanding the logarithm and exponential upto linear order of $(q-1)$ and ignoring terms $\sim [(q-1)^2]$, $X_k$ becomes

\begin{equation}
 X_k=exp(-y_k)F_k~,
 \label{app3}
\end{equation}
with $F_k=1-\frac{1}{2}(1-q)y_k^2$.
Putting (\ref{app3}) into (\ref{app1}) and again expanding the logarithm in an infinite power series we can get,
\begin{eqnarray}
 (1\pm f_q^k)^{q-1}=exp\bigg[(q-1)\sum_{l=1}^{\infty}(\pm)^l\frac{1}{l}e^{-ly_k}F_k^l\bigg]~.
\end{eqnarray} 
By further expansion of the exponential term upto linear in $(q-1)$ we get to Eq.(\ref{iden4}).

Taking its derivative and again keeping terms upto order of $(q-1)$ in $F_k^l$, we finally reach to
the final expression of $\big\{f_q^k\big\}^q$ given by Eq.(\ref{dist5}).
\acknowledgments

I duly acknowledge Vinod Chandra for fruitful discussions and encouragement regarding this project. 
For funding, Indian Institute of Technology Gandhinagar is acknowledged for the Institute Postdoctoral 
Fellowship. I sincerely acknowledge CERN-Theory Division, CERN, Geneva, for hosting the scientific visit in May, 2017,
under CERN visitor program, where a  significant part of the project has been carried out.

\end{document}